# Research on Piano Timbre Transformation System Based on Diffusion Model

Chun-Chieh Hsu[1], Tsai-Ling Hsu[2], Chen-Chen Yeh[3], Shao-Chien Lu[4], Cheng-Han Wu[5], Bing-Ze Liu[6], Timothy K. Shih[7] and Yu-Cheng Lin[8]

[1] Yuan Ze University, Department of Computer Science and Engineering, Taiwan
s1101545@mail.yzu.edu.tw
[2] Yuan Ze University, International Bachelor Program in Informatics, Taiwan
s1093530@mail.yzu.edu.tw
[3] Yuan Ze University, Department of Computer Science and Engineering, Taiwan
yehchenchen93@gmail.com
[4] Yuan Ze University, Department of Computer Science and Engineering, Taiwan
s1111534@mail.yzu.edu.tw
[5] National Central University, Department of Computer Science & Information Engineering, Taiwan,
whoshank@gmail.com
[6] National Central University, Department of Computer Science & Information Engineering, Taiwan,
iojkpghjkhik2@gmail.com
[7] National Central University, Department of Computer Science & Information Engineering, Taiwan,
timothykshih@gmail.com
[8] Yuan Ze University, Department of Computer Science and Engineering, Taiwan
linyu@saturn.yzu.edu.tw

**Abstract.** We propose a timbre conversion model based on the Diffusion architecture designed to precisely translate music played by various instruments into piano versions. The model employs a Pitch Encoder and Loudness Encoder to extract pitch and loudness features of the music, which serve as conditional inputs to the Diffusion Model's decoder, generating high-quality piano timbres. Case analysis results show that the model performs excellently in terms of pitch accuracy and timbral similarity, maintaining stable conversion across different musical styles (classical, jazz, pop) and lengths (from short clips to full pieces). Particularly, the model maintains high sound quality and accuracy even when dealing with rapidly changing notes and complex musical structures, demonstrating good generalization capability. Additionally, the model has the potential for real-time musical conversion and is suitable for live performances and digital music creation tools. Future research will focus on enhancing the handling of loudness dynamics and incorporating additional musical features (such as timbral variations and rhythmic complexity) to improve the model's adaptability and expressiveness. We plan to explore the model's application potential in other timbre conversion tasks, such as converting vocals to instrumental sounds or integration with MIDI digital pianos, further expanding the application scope of the Diffusion-based timbre conversion model in the field of music generation.

## 1    Introduction

Modern composers often integrate various instruments in their albums during the music creation process. However, the recording process for different instruments is complex and time-consuming. Using software sound libraries also requires a high level of skill, as the operation is challenging and the final product often falls short in dynamic performance and sound quality.

   The creation and use of piano timbres are equally challenging. Piano timbres are nuanced, and their varied playing techniques require careful consideration in the production and use of piano sounds. In terms of timbre, pianists can adjust the richness and emotional expression by controlling key strike intensity, pedal use, and speed variation, allowing piano timbres to range from gentle and delicate to powerful and grand, displaying high musical expressiveness. In terms of playing techniques, the piano includes various expressive methods, such as tremolo, glissando, and chord arpeggios. These techniques enable the piano to convey rich emotional layers and perform complex technical expressions. For instance, pedal techniques can prolong the duration of notes, creating richer sound effects, while controlling the force and speed of key strikes can enhance the layers of the melodic line. However, due to the delicacy of piano timbres and the variety of techniques, commercially available piano sound sources often fail to achieve the effect of a real piano performance.

   Therefore, to lower the barriers to using piano timbre conversion and to convert target audio into sounds closer to a real piano, we aim to use deep learning methods to transform target music into realistic piano timbre versions, enhancing the quality of musical works. To this end, we have designed an autoencoder architecture for timbre conversion. First, we convert the input target music into a generic representation that matches its features, while reducing its dimensions. Then, these low-dimensional features are used as generative conditions in an unconditional diffusion model, which converts them into corresponding piano sounds.  Finally, by using the trained model, users can input any .wav file and obtain a .wav file converted into piano sounds, achieving excellent piano timbre performance without the need for specialized knowledge and skills.

## 2    Related work

Piano timbre conversion is a technique that transforms the unique sound characteristics of the piano into sounds similar to other instruments; with advancements in deep learning, this technology has gained widespread attention and application in the field of music technology. Recurrent Neural Networks (RNNs) and Long Short-Term Memory networks (LSTMs) play a key role in piano timbre conversion; these models excel at processing the temporal sequence data of piano audio, capturing the complex characteristics of piano sound over time, and preserving its details and texture during the conversion process. For example, LSTM models excel at capturing long-term dependencies, learning the temporal dynamic characteristics of piano timbres, including note sustenance and dynamic changes, as presented by Qian Liu[1] (2023) in a study on deep learning-based piano music aesthetic emotional expression. Furthermore, Gated Recurrent Units (GRUs), a variant of RNNs, further enhance the efficiency and effectiveness of processing longer time sequences, making the timbre conversion results more coherent and natural, as proposed by M. Sahaya Sheela, S. Surendran, P. Muthu Krishnammal, Vinothkumar K, Surendra Shetty, B. Venkataramanaiah[2] (2023) in improvements to AI-based piano music recognition algorithms. Autoencoders, especially Variational Autoencoders (VAEs), encode piano audio into a low-dimensional latent space and decode it back to audio data, achieving timbre conversion while preserving the core features of piano timbres, as proposed by Shih-Lun Wu, Yi-Hsuan Yang [3] (2023) in MuseMorphose: using VAEs for whole-piece and fine-grained piano music style conversion.

Generative Adversarial Networks (GANs) offer an innovative approach to achieve high-fidelity piano timbre conversion. GANs use adversarial training between generators and discriminators to make generated piano audio samples more realistic and natural, especially suitable for situations that do not require paired training data. This technique allows for the conversion between piano timbres and other instrument timbres within a bidirectional learning framework, as proposed by Lijun Zheng, Chenglong Li[4] (2024) in real-time emotional piano music generation based on GANs. Additionally, GANs can handle the conversion of multiple instrument timbres simultaneously through multi-domain conversion technologies, greatly enhancing the model's efficiency and scalability, as proposed by Yigitcan Özer, Meinard Müller[5] (2024) in using musical motif enhancement techniques to separate piano concerto sources, and by Wenxuan Jiang, Yaqi Guo, Jiulian Li[6] (2023) in a method for digital piano audio signal feature identification based on convolutional neural networks. The flexibility of GANs has become a powerful tool in piano timbre conversion and can be combined with other technologies to further enhance the fidelity (High Fidelity, Hi-Fi) and granularity of the timbre conversion.

Future piano timbre conversion technologies will focus on higher authenticity, smoother transition processes, and broader application scenarios. Researchers are dedicated to exploring new neural network architectures and deep learning algorithms to improve the precision and efficiency of timbre conversion. Deep reinforcement learning has been applied in piano fingering generation, as proposed by Ananda Phan Iman, Chang Wook Ahn[7] (2024) in a model-free deep reinforcement learning method for piano fingering generation and polyphonic music transcription systems. Taehyeon Kim, Donghyeon Lee, Man-Je Kim, Chang Wook Ahn[8] (2024) proposed a system for polyphonic piano music transcription leveraging inter-note relationships. These applications demonstrate the potential and direction of timbre conversion technology. Additionally, technologies based on time-frequency information and convolutional neural networks are being developed to improve piano music recommendation algorithms, as proposed by Sivagami. S, S. Ramani, Vishal Mehra, L. H. Jasim, M. Sahaya Sheela,

Kavitha S[9] (2023) in a study on time-frequency-based piano music recommendation algorithms and feature identification methods, and by Andres Fernandez[10] (2023) in real-time piano transcription using convolutional neural networks. As technology continues to evolve, timbre conversion technology is expected to play a greater role in fields such as music creation, education, and entertainment, providing more possibilities for musical expression and creation, as proposed by Ruimin Wu, Xianke Wang, Yuqing Li, Wei Xu, Wenqing Cheng[11] (2024) in piano transcription with harmonic attention.

## 3   Related Technique

Our key technologies used in generative models include Diffusion Models, V-Diffusion, CREPE, and U-Net, with a focus on their applications in data generation and processing.

### 3.1   Diffusion model

Denoising Diffusion Probabilistic Models (DDPM). DDPM is a generative model that reconstructs data by progressively removing noise. At its core, it treats the generation of data as a reverse process, gradually restoring the original data from random noise. The generative process of DDPM consists of a series of reverse Markov processes, where the model learns to produce realistic data samples from Gaussian noise. Compared to other generative models, DDPM exhibits significant advantages in terms of training stability and the quality of generated samples.

Denoising Diffusion Implicit Models(DDIM). DDIM is an improvement over the Denoising Diffusion Probabilistic Model (DDPM), reducing computational burden through a deterministic generation process. DDIM decreases the number of steps required for generation and enhances efficiency, making it suitable for scenarios that demand rapid generation of high-quality data.

### 3.2   V-diffusion

V-Diffusion is a diffusion model focused on music generation, which simulates the distribution of music by progressively adding and removing noise, thereby reconstructing high-quality music segments. The innovation of this model lies in its use of a learnable noise schedule, enhancing musical details and sound quality.

### 3.3   CREPE

CREPE is a one-dimensional convolutional neural network designed for pitch estimation. By processing audio signals, this model can automatically extract pitch-related features from raw audio data, demonstrating excellent performance in music analysis.

### 3.4 U-Net

U-Net is a neural network designed for image segmentation, which extracts and restores image features through a contracting path (Encoder) and an expansive path (Decoder). The skip connection technique of this model preserves spatial information of the image, enabling precise image segmentation.

## 4 Method Architecture

In this paper, we present the architecture of the model in the following **Fig. 1** and **Fig. 2** where the methods mentioned below will be introduced.

Pitch is a crucial element in music composition, and ensuring the accuracy of pitch over time is vital for the effectiveness of timbre conversion. To achieve this, we use pitch information as an input condition to guide the generative process of the diffusion model. The **pitch encoder** takes the target audio file that needs to be converted as input. We first use the CREPE pitch tracker to determine the fundamental frequency of the input audio waveform. Next, the estimated fundamental frequency is tokenized and discretized, which helps the model accurately capture the corresponding pitch in the audio. After converting the audio waveform into the corresponding pitch, we one-hot encode the results and transform them into a more complex vector representation through a fully connected layer. This process enhances the model's effectiveness in subsequent learning.

This method ensures the accuracy and effective application of pitch information, preserving the structure and sound quality of the music during model training and the conversion process. By using pitch as a key controlling factor, we maintain pitch consistency between the converted and original audio, thereby ensuring the high quality and reliability of the conversion results.

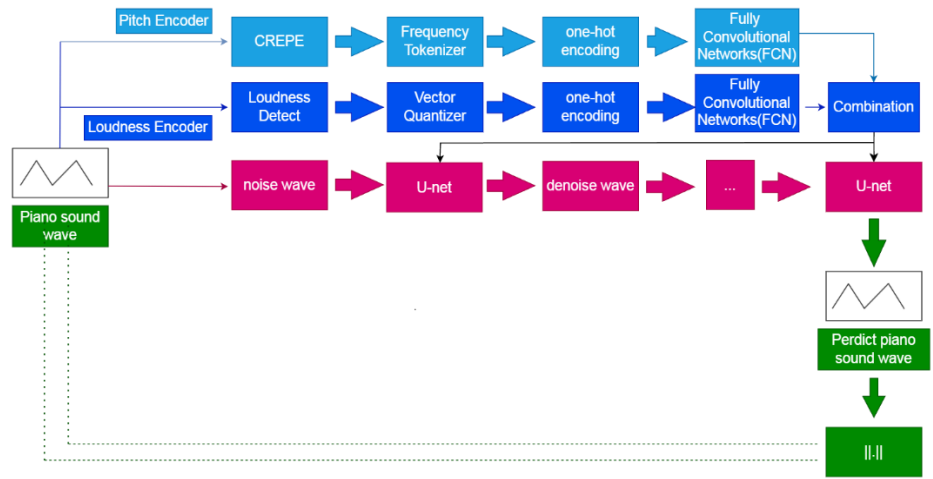

**Fig. 1.** The flowchart of model training process.

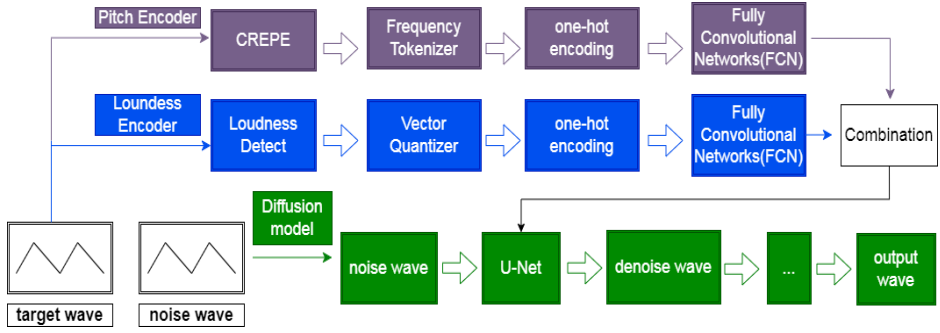

**Fig. 2.** The complete architecture of the model.

Embedding is a fundamental concept in natural language processing (NLP). In NLP, embedding refers to the mapping of words or phrases into a lower-dimensional vector space. These vectors capture specific structural and semantic relationships, enabling the model to identify similarities and connections be-tween words. By leveraging embedding, we can transform text into a format that machines can interpret and process, applying these vector representations to various NLP tasks. The theoretical foundation of embedding is the Distributional Hypothesis, which states that words with similar meanings tend to appear in similar contexts. Based on this hypothesis, embedding models learn to map semantically related words to adjacent positions in the vector space, allowing for the effective calculation and comparison of distances or similarities between them.

In our timbre conversion model, embedding also plays a crucial role. We convert the pitch indices obtained from the frequency tokenizer into corresponding one-hot vectors. These one-hot vectors are high-dimensional and sparse, with only one element set to 1 and the rest to 0. To reduce the dimensionality of these high-dimensional sparse vectors generated by one-hot encoding, we apply a fully connected layer to transform them into low-dimensional dense vectors, which are the learned embedding. These dense vectors more effectively reflect the similarities and differences between pitches, while improving computational efficiency and scalability.

During the timbre conversion process, the use of pitch indices is critical. We obtain the pitch indices from the frequency tokenizer and convert them into corresponding vector forms. Table 1. illustrates the pitch indices, their corresponding pitch names, and fundamental frequencies (F0, in Hz), providing the pitch information ranging from C4 to B6, along with their respective indices and fundamental frequency values.

**Table 1.** Fundamental Frequency Interval Reference Table

| Index | Pitch | F0(Hz) | Index | Pitch | F0(Hz) |
|---|---|---|---|---|---|
| 0 | None | 0 | 19 | F#5/ G♭5 | 739.99 |
| 1 | C4 | 261.63 | 20 | G5 | 783.99 |
| 2 | C#4/ D♭4 | 277.18 | 21 | G#5/ A♭5 | 830.61 |
| 3 | D4 | 293.66 | 22 | A5 | 880.00 |
| 4 | D#4/ E♭4 | 311.13 | 23 | A#5/ B♭5 | 932.33 |
| 5 | E4 | 329.63 | 24 | B5 | 987.77 |
| 6 | F4 | 349.23 | 25 | C6 | 1046.50 |
| 7 | F#4/ G♭4 | 369.99 | 26 | C#6/ D♭6 | 1108.73 |
| 8 | G4 | 392.00 | 27 | D6 | 1174.66 |
| 9 | G#4/ A♭4 | 415.30 | 28 | D#6/ E♭6 | 1244.51 |
| 10 | A4 | 440.00 | 29 | E6 | 1318.51 |
| 11 | A#4/ B♭4 | 466.16 | 30 | F6 | 1396.91 |
| 12 | B4 | 493.88 | 31 | F#6/ G♭6 | 1479.98 |
| 13 | C5 | 523.25 | 32 | G6 | 1567.98 |
| 14 | C#5/ D♭5 | 554.37 | 33 | G#6/ A♭6 | 1661.22 |
| 15 | D5 | 587.33 | 34 | A6 | 1760.00 |
| 16 | D#5/ E♭5 | 622.25 | 35 | A#6/ B♭6 | 1864.66 |
| 17 | E5 | 659.25 | 36 | B6 | 1975.53 |
| 18 | F5 | 698.46 | | | |

**Frequency Tokenizer** plays a crucial role in natural language processing, especially in the field of deep learning, where text needs to be converted into numerical form to be processed by neural networks. The primary purpose of a frequency tokenizer is to split the input text into smaller units, such as words, subwords, or characters, and map these units to specific indices or word vectors. This process transforms text strings into variable-length one-dimensional sequences that the model can understand.

In our approach, we use the fundamental frequency predicted by CREPE as input and convert it into corresponding pitch values. Pitch is a perceptual attribute that determines how high or low a sound is perceived by the human ear, which is directly related to the frequency of sound waves. Generally, higher frequencies correspond to higher pitches, while lower frequencies correspond to lower pitches. Since pitch more intuitively represents how humans perceive sound compared to fundamental frequency, we use a pitch frequency table to convert the fundamental frequency into corresponding pitch values. This reduces the computational burden on the model, allowing it to focus more on capturing the rhythmic patterns of the input waveform and accurately reproducing the unique timbre of the piano.

Considering the typical pitch range of piano music spans several octaves, we focus on the most commonly used range in our dataset. We classify the predicted fundamental frequencies based on their proximity to pitch frequencies within this specific range. These classified groups are then converted into indices, arranged in order from low to high pitch. This classification method enables the model to more effectively integrate pitch information into the conversion process, improving the overall quality and accuracy of the results. Additionally, we set the fundamental frequency of 0 to index 0, indicating silent or non-expressive parts of the piano waveform, further refining the design of the conversion system.

Besides pitch, loudness is also an important factor affecting piano performance. Loudness describes the volume of sound perceived by the human ear, which is a physical quantity related to the sound intensity of sound waves. However, due to the human ear's different sensitivities to different frequencies, the perception of loudness varies with frequency even at the same sound intensity. To more accurately reflect human auditory perception, we chose loudness as a control variable in our model.

In the **Loudness Encoder**, we input the target audio file to be converted. To reduce computational load, we first divide the input piano sound wave into 16 equal-length segments. This segmentation method not only reduces computational complexity but also effectively captures the loudness characteristics of the piano sound wave over time. By doing so, we can retain the subtle changes in the sound wave at each time point, which is crucial for subsequent feature extraction and analysis. We then measure the loudness of each segment according to the ITU-R BS.1770-4 standard, setting the minimum loudness value at -70 LKFS. Unlike pitch, loudness cannot be directly represented or quantified like musical notes. Therefore, we use a Vector Quantization Model to identify the most common loudness values in the audio and convert these values into indices, thereby achieving further data compression and feature simplification.

To effectively integrate with pitch features, we scale these indexed loudness features proportionally so that their length on the time axis matches the pitch features extracted by the Pitch Encoder, maintaining consistency between loudness and pitch features. This enhances the model's effectiveness in learning these features. Similarly, these compressed indexed loudness data are input into the embedding model to generate more expressive feature embedding, helping the model achieve higher quality in generating piano tones.

In the task of timbre conversion, accurately representing loudness is crucial for maintaining the dynamic range and emotional expression of music. However, continuous loudness values can increase the computational burden on the model and adversely affect the training and convergence speed. To overcome these issues, we employed a VQ, an efficient data compression technique that converts continuous loudness data vectors into discrete codes from a finite set, achieving compression and encoding of loudness data.

The operation principle of the **Vector Quantizer** involves clustering the loudness vectors in the training data to generate a representative codebook. This codebook contains representative information of various loudness features, effectively reflecting the distribution characteristics and statistical properties of loudness. During the encoding process, the input loudness vector is mapped to the closest codeword in the codebook and represented using the corresponding code. This way, we can transform continuous loudness data into discrete encoded forms, not only achieving data compression and encoding but also effectively reducing storage and transmission costs.

In the process of reconstructing an upsampled structure similar to the original input from encoded vectors, we drew inspiration from Flavio Schneider's architectural[12] design. This design suggests using a diffusion model as a decoder to generate high-quality piano timbres. By employing U-Net as the backbone of the diffusion model, we naturally integrated the autoencoder functionality with skip connections. The latent vectors generated by the encoder are incorporated into the appropriate levels of the U-Net, effectively serving as the "decoder" in this framework. Unlike the original design, which removed the self-attention mechanism in the U-Net to accommodate different waveform lengths, we chose to maintain fixed input and output waveform lengths while retaining the self-attention mechanism in the U-Net.

This approach enables the model to accurately reconstruct the temporal relationships of features extracted by the pitch and loudness encoders, allowing it to focus on critical musical attributes while reducing reliance on irrelevant information. The self-attention mechanism allows the model to dynamically adjust the focus at each time step during the generation process, effectively capturing key details in the musical piece. This flexibility enables the model to adapt to different styles and complexities of music, thereby enhancing the accuracy and naturalness of the generated piano timbres.

The outputs of the pitch and loudness encoders are combined and used as conditional inputs to the model. The combined conditional inputs are fed into the downsampling layers of the U-Net, providing the necessary features to generate piano timbres with consistent pitch and loudness.

## 5    Case Analysis

### 5.1    Dataset

Dataset of case comprises 159 pieces of Jiangnan style of traditional Chinese music in 4/4 time, with a total duration of approximately 6919 seconds. These pieces are based on the pentatonic scale, lacking the notes Fa and Si. The music tracks were recorded at a sampling rate of 44100 Hz and segmented into 6-second intervals. The following figure and tables (see **Fig. 3**, **Table 2**, **Table 3**) respectively present the pitch and beat length information within the dataset.

In **Table 3**, under the scenario of a 4/4 time signature with a tempo of 60 beats per minute, each measure lasts 2 seconds, and each quarter note lasts 0.5 seconds. However, for convenience in the program, time is usually represented in ticks. In this study, we define the duration of a quarter note as a fixed 480 ticks, making each measure 1920 ticks and each second 960 ticks. This value remains constant regardless of tempo changes.

In traditional Jiangnan music, the rhythmic structure prominently features quarter notes and eighth notes. Given this focus on quarter and eighth notes, the full range of 1920 available tick values is unnecessary for accurate representation. Instead, we focus on just 10 specific tick values that are most frequently observed in the rhythmic patterns of Jiangnan music. By concentrating on these 10 tick values, we streamline the analysis and ensure that our model accurately reflects the rhythmic tendencies of Jiangnan music without being overwhelmed by extraneous tick values that do not play a substantial role in its structure.

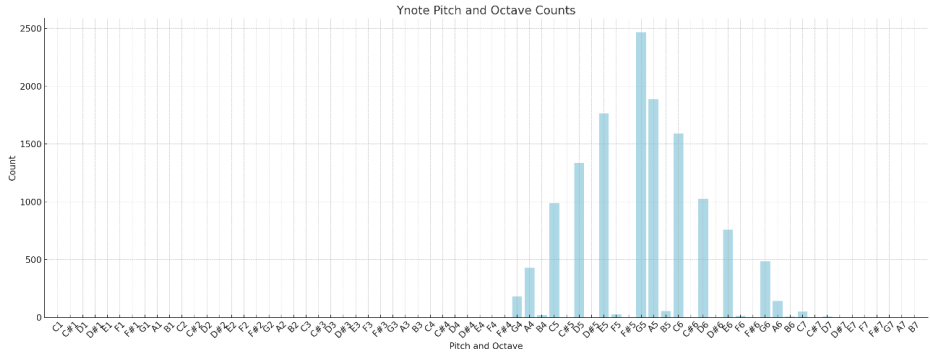

**Fig. 3.** Pitch Statistics of the Dataset (C1 to B7)

**Table 2.** Pitch Statistics of the Dataset (C1 to B7)

| Pitch_0ct | Count | Pitch_0ct | Count | Pitch_0ct | Count |
|---|---|---|---|---|---|
| E4 | 2 | F5 | 24 | F6 | 10 |
| G4 | 178 | G5 | 2465 | G6 | 485 |
| A4 | 429 | A5 | 1887 | A6 | 141 |
| B4 | 20 | B5 | 54 | B6 | 4 |
| C5 | 987 | C6 | 1591 | C7 | 49 |
| D5 | 1337 | D6 | 1027 | D7 | 5 |
| E5 | 1763 | E6 | 758 | | |

**Table 3.** Statistical Analysis of Note Lengths in the Dataset

| Tick | Note Length | Count |
|---|---|---|
| 180 | Dotted Sixteenth Note | 75 |
| 240 | Eighth Note | 4813 |
| 360 | Dotted Eighth Note | 531 |
| 480 | Quarter Note | 1017 |
| 720 | Dotted Quarter Note | 209 |
| 960 | Half Note | 572 |
| 1440 | Dotted Half Note | 20 |
| 1920 | Whole Note | 16 |
| 3840 | Double Whole Note | 70 |
| 160 | Quarter Note Triplet | 8 |

### 5.2   Training

We randomly selected segments from the original waveforms of 159 training samples, with each segment consisting of $2^{18}$ frames, to serve as input for the model. The training was conducted on an NVIDIA RTX 4090 GPU with a batch size of 8 over 6000 epochs, taking a total of 4 days.

### 5.3   Evaluation

The following discussion is divided into four situations:

1. The pitch is within the range of **Table 2** and the note length is within the range of **Table 3**

   The left side of **Fig. 4** shows the spectrum of the original recording of Twinkle, Twinkle, Little Star played with a violin, while the right side of **Fig. 4** shows the spectrum of the piano tone version generated by the model using the violin playing as input. In **Fig. 5**, a comparison of two spectra in the audible frequency range shows that their spectral characteristics are highly similar. This shows that the model successfully produces a piano timbre that is very similar to the sound of a real piano, while maintaining similar frequency distribution characteristics to the original violin version. These results show that the model performs effectively within the specified pitch range (see **Table 2**) and note duration range (see **Table 3**).
   **Example_1:** *Twinkle, Twinkle, Little Star* played by violin.

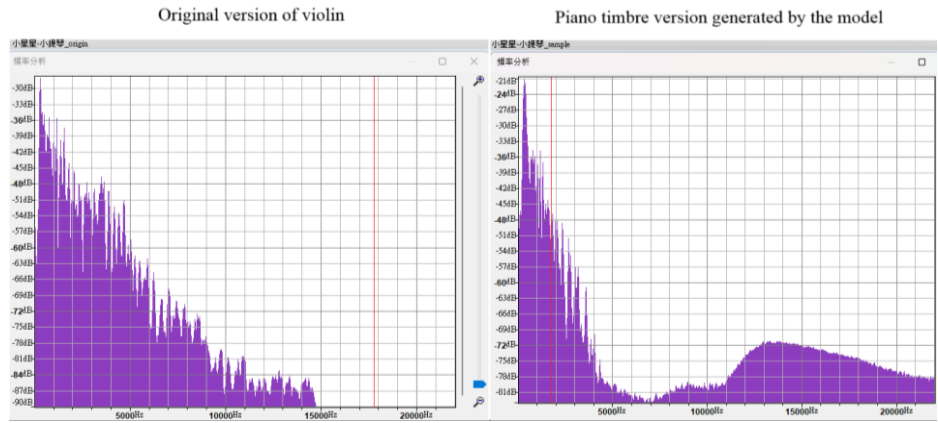

**Fig. 4.** The left side is the loudness distribution of *Twinkle, Twinkle, Little Star* played with a violin where the fundamental frequency and harmonic peaks are shown, while the right side is the model-generated piano timbre using the violin playing as input.

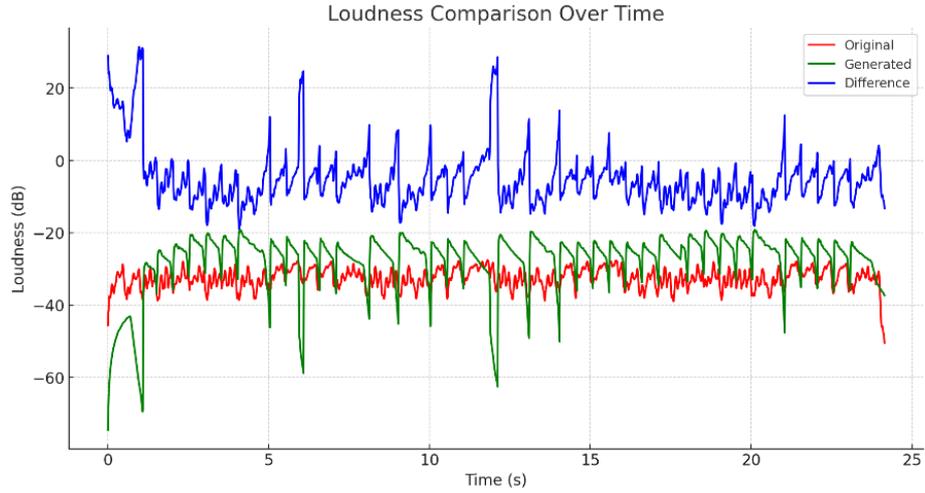

**Fig. 5.** The loudness variation between the original violin audio (red line), the piano timbre version (green line), and their difference (blue line) over time, revealing the temporal loudness discrepancies between the two versions.

We explored different music and instruments as input for the model, such as Example 2, where *Taiho Boat* is played by the flute, and Example 3, where *Jasmine* is played by the bass. The model then generated piano versions of these songs. As observed in **Fig. 7** and **Fig. 9**, the difference in loudness over time and spectral characteristics between the original and generated versions is minimal. This indicates that the spectral properties of the generated piano version are highly similar to the original audio of the actual instruments. These results further suggest that the model successfully synthesized piano sounds within the specified pitch range (see **Table 2**) and note duration range (see **Table 3**) that closely resemble real piano audio while maintaining a similar frequency distribution to the original instrument versions, thus demonstrating the model's effectiveness.

**Example_2:** *Taiho Boat* is played by flute.

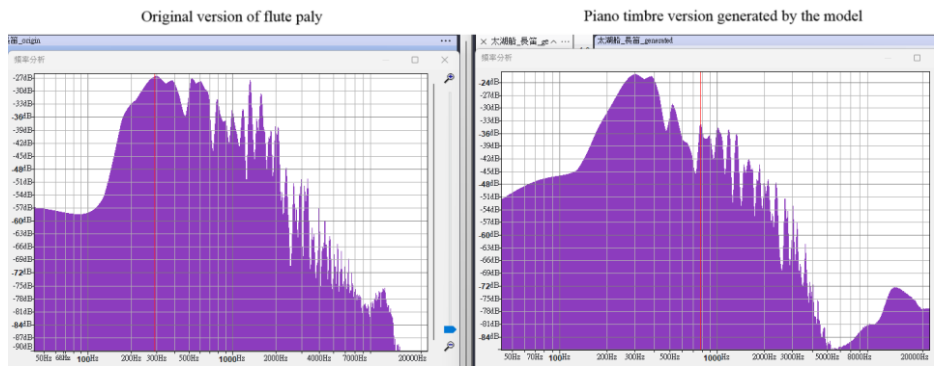

**Fig. 6.** The right side is original version played by flute and the left side is the model-generated piano timbre using the flute playing as input.

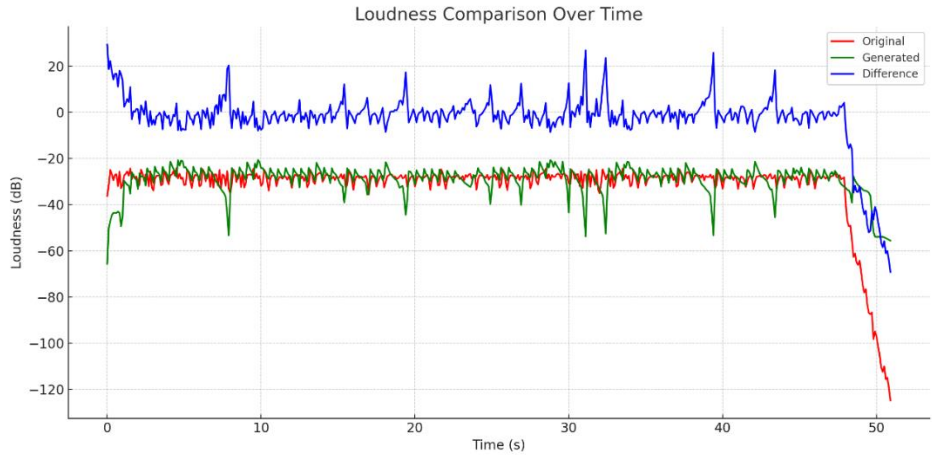

**Fig. 7.** The loudness variation between the original flute audio (red line), the generated piano timbre version (green line), and their difference (blue line) over time reveals that the temporal loudness discrepancies between the two versions are similar.

**Example_3:** *Jasmine* is played by bass.

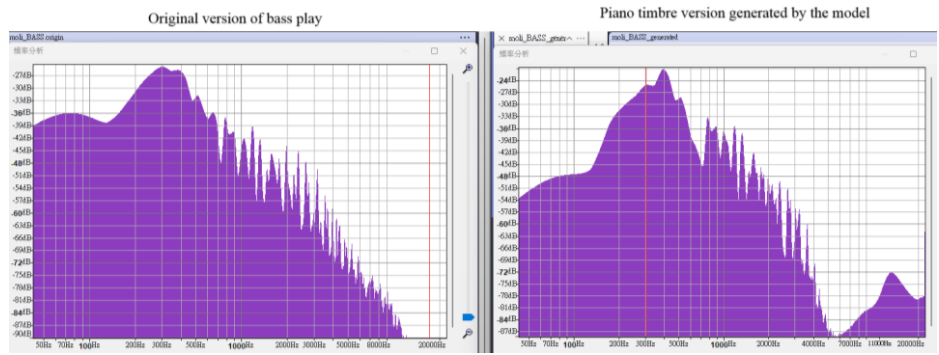

**Fig. 8.** The right side is original version played by bass and the left side is the model-generated piano timbre using the bass playing as input.

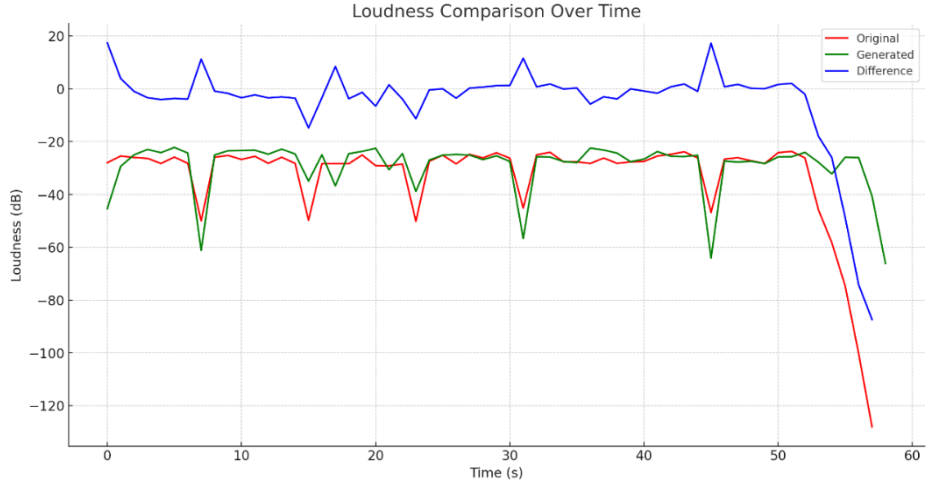

**Fig. 9.** The loudness variation between the original flute audio (red line), the generated piano timbre version (green line), and their difference (blue line) over time indicates that the spectral characteristics of the two versions are highly similar.

2. One Octave Lower Pitch and the note length is within the range of **Table 3**:
   The left side of **Fig. 10** shows the original version of *Twinkle, Twinkle, Little Star* performed on the violin one octave lower, while the right side of **Fig. 10** presents the piano timbre version generated by the model using the lower octave version as input. A comparison of the two spectra in **Fig. 11** reveals significant differences within the audible frequency range. Although the pitch is lowered by one octave and the note durations remain the same, the model-generated piano timbre exhibits insufficient energy in the lower frequency range when compared to the original violin version. This discrepancy could be attributed to a relative lack of training data in the C1 to C3 octave range.

   Based on the data analysis from **Fig. 3** and **Table 2**, the distribution of training data significantly influenced the model's ability to generate piano timbres, particularly in the lower frequency region. The scarcity of data in this range impaired the model's ability to accurately reproduce timbres in the bass region. Consequently, this difference highlights the critical impact of training data distribution on the model's generation performance, especially when recreating timbres in the lower register.

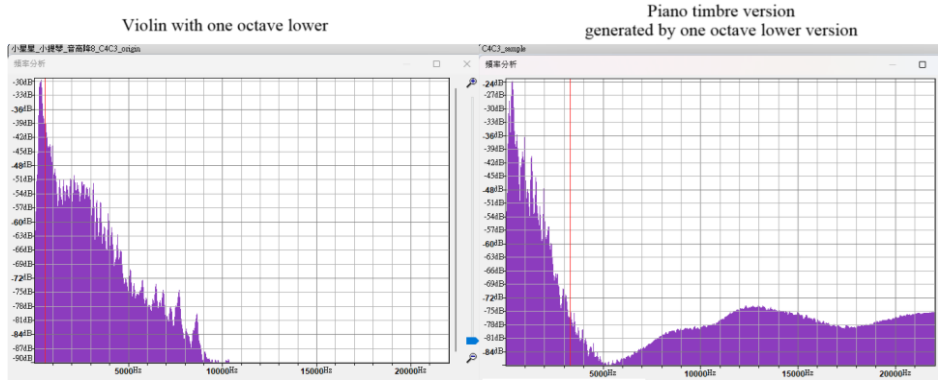

**Fig. 10.** The left side illustrates the loudness distribution of the violin's lower octave version, while the right side displays the loudness distribution of the model-generated piano timbre.

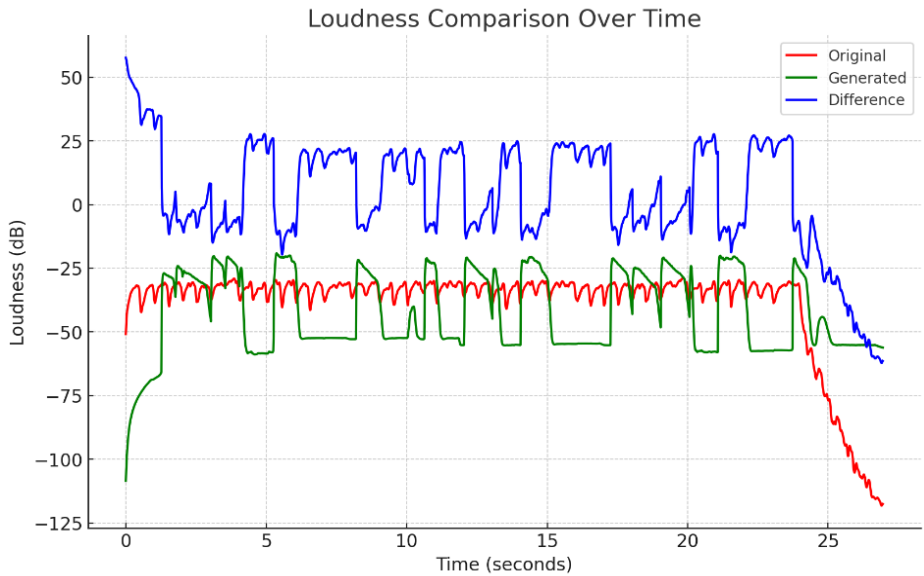

**Fig. 11.** Comparing the loudness variations between the original and generated audio, revealing pronounced differences in the low-frequency range. These charts underscore the importance of the training data's distribution in the process of timbre generation, especially when dealing with low-frequency regions.

3. The pitch is within the range of **Table 2** and the note length is Eighth Note Duration:
   **Fig. 12** respectively display the violin performance and the piano version of Twinkle, Twinkle, Little Star generated by the model. Both versions are based on the same pitch for eighth notes. From **Fig. 13**, we can observe that the two versions show a highly similar distribution within the audible frequency range. This indicates that the transformation model effectively preserves the timbral characteristics of the original music segment and successfully simulates the piano timbre. The similarity shown in **Fig. 13** further confirms the model's reliability and effectiveness in timbre conversion between different instruments.

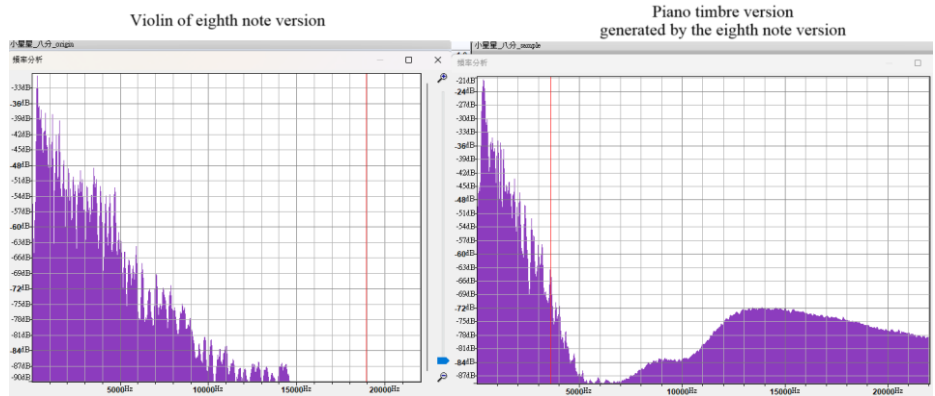

**Fig. 12.** Respectively display the violin performance and the piano version of *Twinkle, Twinkle, Little Star* generated by the model.

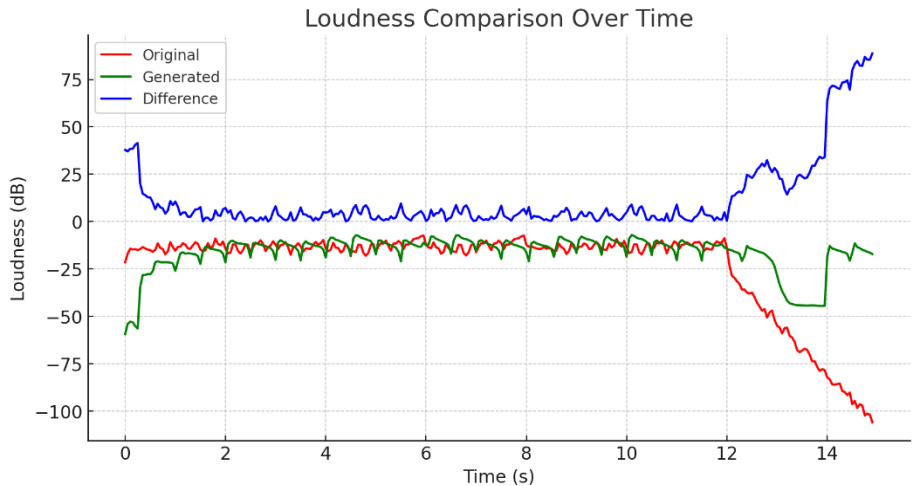

**Fig. 13.** A comparison of the loudness between the original violin version (red line) and the generated piano version (green line) and the blue shows the difference between the two. The x-axis represents time (seconds), and the y-axis represents loudness (dB). The loudness variation over time is highly similar between the two versions, with only minor deviations.

4. One Octave Lower Pitch and the note length is Thirty-second Note Duration:
    The left side of **Fig. 14** displays the original version of *Twinkle, Twinkle, Little Star* performed on the violin, presented in a lower octave with a thirty-second note

rhythm. The spectrum shows that energy is concentrated predominantly in the lower frequency range, particularly below 1000 Hz, highlighting strong low-frequency characteristics. The right side of **Fig. 14** illustrates the piano version generated by the model. The spectrum reveals a broader frequency distribution, with relatively stable energy responses in the higher frequency range, indicating a wider spread of energy across the frequency spectrum in the generated piano version. **Fig. 15** shows the loudness difference highlights significant discrepancies between the two versions, particularly in their dynamic range.

These differences could be attributed to imperfections in the training dataset. According to the analysis in **Fig. 3** and **Table 2**, the training data lacked sufficient samples in the C1 to C3 range, leading to weaker performance in the low-frequency band of the generated version. Additionally, the dataset did not include examples of thirty-second note rhythms, further affecting the accuracy of the generated music in terms of rhythm and temporal characteristics.

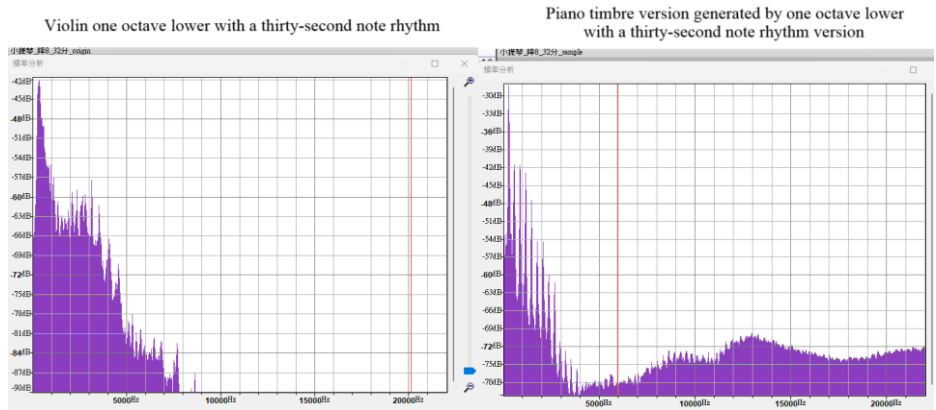

**Fig. 14.** The left side displays the original version performed on the violin, presented in a lower octave with a thirty-second rhythm, and right side illustrates the piano version generated by the model.

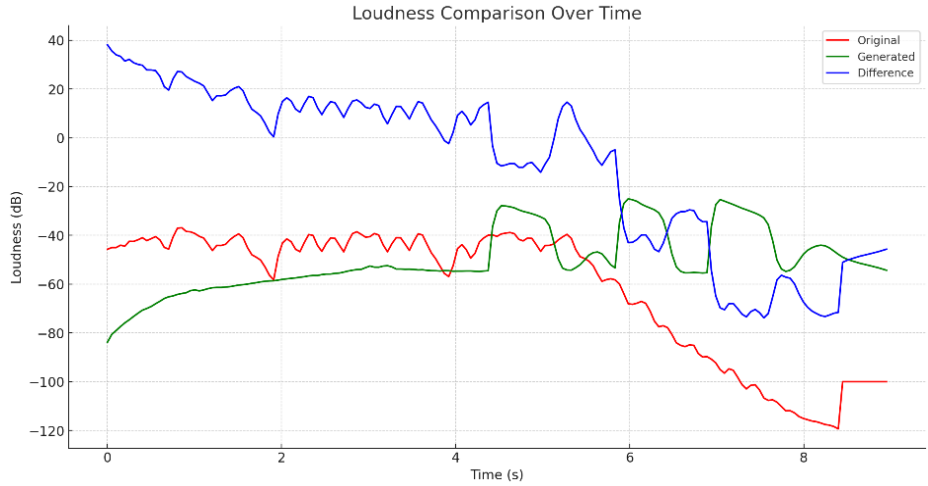

**Fig. 15.** Comparing the loudness differences (blue line) between the original (red line) and generated versions (green line).

## 6     Conclusion

In this paper, we propose a timbre conversion model based on a diffusion architecture that transforms any music into piano timbre. The model first extracts common features from musical performances and then inputs them into the diffusion model decoder to reconstruct the corresponding piano performance. In the project analysis phase, the initial training dataset consisted of traditional Chinese Jiangnan-style music, which is primarily based on the pentatonic scale and lacks the pitches Fa and Si. This resulted in an incomplete dataset, which could have influenced the model's performance. To address this limitation, we designed multiple tests to evaluate the model's performance, with a particular focus on pitch accuracy. The results demonstrate that the model performed well in maintaining pitch accuracy and was able to generate sounds closely resembling the timbre of a real piano. Despite the incomplete dataset, the model effectively captured the essential characteristics of the input music and successfully translated them into a coherent piano output.

Research on Piano Timbre Transformation System    19